\documentclass[twocolumn,preprintnumbers,amsmath,amssymb]{revtex4}
\usepackage{graphicx}
\usepackage{float}
\usepackage[usenames,dvipsnames]{color}
\newcommand{\req}[1]{Eq.\,({\ref{#1}})} 
\begin{document}
%\preprint{Draft.13}
\title{Possibility of bottom-catalyzed matter genesis near to primordial QGP hadronization}
\author{Cheng Tao Yang${}^a$, and Johann Rafelski${}^a$}
\affiliation{${}^a$Department of Physics, The University of Arizona, Tucson, Arizona 85721, USA}
\date{April 14, 2020}

%%%%%%%%%%%%%%%%%%%%%%%%%%%%%%%%%%%%%%%%%%%%%%%%%%%%%%%%%%%%%%%%%%
\begin{abstract}
We study bottom flavor abundance in the early Universe near to a temperature $T_\mathrm{H}\simeq150\,\mathrm{MeV}$, the condition for hadronization of deconfined quark-gluon plasma (QGP). We show bottom flavor abundance nonequilibrium lasting microseconds. In our study we use that in both QGP, and the hadronic gas phase (HG) $b$ and $\bar b$ quarks near $T_\mathrm{H}$ are bound in B-mesons and antimesons subject to $CP$ violating weak decays. A coincident non-equilibrium abundance of bottom flavor can lead to matter genesis at required strength: a) The specific thermal yield per entropy is $n_b^{th}/\sigma=10^{-10}\sim10^{-13}$. b) Considering time scales, millions of cycles of  B-meson decays, and $b\bar b$-pair recreation processes occur.
 \end{abstract}
\maketitle

%%%%%%%%%%%%%%%%%%%%%%%%%%%%%%%%%%%%%%%%%%%%%%%%%%%%%%%%%%%%%%%%%%%%%%%%%%%%%%%%%%%%
\section{Introduction} 
\subsection{Overview}
An epoch in the primordial Universe evolution allowing matter genesis (baryogenesis, leptogenesis) at the level observed has not been established. Usually the temperature range between GUT phase transition $T_\mathrm{G}\simeq10^{16}\,\mathrm{GeV}$ and the electroweak phase transition near $T_\mathrm{W}\simeq130\,\mathrm{GeV}$ is explored~\cite{Kuzmin:1985mm,Kuzmin:1987wn,Arnold:1987mh,Kolb:1996jt,Riotto:1999yt,Nielsen:2001fy,Giudice:2003jh,Davidson:2008bu,Morrissey:2012db}. Here we present arguments that the Sakharov conditions~\cite{Sakharov:1967dj} for matter asymmetry to form also appear during quark-gluon plasma (QGP) hadronization era near to $T_\mathrm{H}\simeq150\,\mathrm{MeV}$. We show substantial departure from equilibrium and $C$ and $CP$ violation. 

The third condition, a violation of baryon $B$, lepton $L$ number conservation coincident with the above situation needs future theoretical and experimental consideration, constrained by the experimental limit on  proton life span $\mathcal{O}(10^{32}\mathrm{y}$). The relevant thermal environment can be explored in the laboratory since $T_\mathrm{H}$ is readily available in relativistic heavy ion (RHI) collision experiments~\cite{Rafelski:2019twp}. However, $T_\mathrm{H}$  may not suffice for catalysis of baryogenesis~\cite{Kuzmin:1985mm,Kuzmin:1987wn,Arnold:1987mh}. On the other hand, if leptoquarks exist~\cite{Bauer:2015knc}, it is possible that the decay of heavy bottomnium flavored mesons could generate, via $B-L=0$ processes, the matter excess required. We defer this question reaching beyond the scope of this work to future studies.

The observed baryon and lepton number cannot be generated in a full thermal (chemical and kinetic) equilibrium, because even if the required processes are occurring, the net effect is cancelled out by the equal number of back-reactions. We believe that the presence of abundance ({\it i.e.\/} chemical) non-equilibrium is more relevant -- kinetic (equipartition of energy) equilibrium is usually established much quicker and has less impact on the actual particle abundances~\cite{Koch:1986ud,Birrell:2014gea}. In this work we show that a relatively large bottom flavor chemical non-equilibrium arises near to the QGP hadronization condition.

The Sakharov condition requiring $C$ and $CP$ assures that we can recognize a universal difference between matter and antimatter, thus one abundance can be enhanced compared to the other. We are seeking $CP$ violation of relevance to QGP,  considering a mechanism grossly different and presumably entirely independent from the chiral magnetic effect~\cite{Kharzeev:2007jp}. Specifically, given that the non-equilibrium of bottom flavor arises at relatively low QGP temperature, the bottom quark decay occurs from preformed~\cite{Karsch:1987pv,Brambilla:2010vq,Aarts:2011sm,Brambilla:2017zei,Bazavov:2018wmo,Offler:2019eij} $\mathrm{B}_x$ meson states, $x=u,d,s,c$. These decays violate aside of $C$ also the $CP$ symmetry, see for example~\cite{Aaij:2019hzr,Aaij:2020alb}. The exploration of the here interesting $CP$ symmetry breaking in B$_c(b\bar c)$ decay is in progress~\cite{Tully:2019ltb,Amhis:2019ckw}. While in the following we focus our attention on the QGP deconfined phase, in qualitative and nearly quantitative manner our results apply to bottomnium non-equilibrium and $CP$ violation in the hadron gas (HG) phase.

The $CP$ violation is well established in all bottom mesons including B$_c(b\bar c)$ decays~\cite{Tanabashi:2018oca}. In general, violation of $CP$ asymmetry can occur in the amplitudes of hadron decay. The weak interaction $CP$ violation arises from the components of Cabibbo-Kobayashi-Maskawa (CKM) matrix associated with quark-level transition amplitude and $CP$-violating phase. In this case, the charged B$_c$ meson decay can be the source of required $CP$ violation~\cite{Aaij:2019hzr}.

\subsection{Is there enough bottom flavor to matter?} 
Considering that the expanding Universe evolves conserving entropy, and that baryon and lepton number following on the era of matter genesis is conserved, the current day baryon $B$ to entropy $S$, $B/S$-ratio must be achieved during matter genesis. The PDG~\cite{Tanabashi:2018oca} estimates the present day baryon-to-photon ratio $5.8 \times 10^{-10} \leqslant\eta\leqslant6.5\times10^{-10}$. This small value quantifies the matter-antimatter asymmetry in the present day Universe. The parameter $\eta$ allows the determination of the present value of $B/S\approx7.69\times10^{-11}$~\cite{Rafelski:2019twp,Letessier:2002gp,Fromerth:2002wb,Fromerth:2012fe} in the Universe dominated by photons and free-streaming low mass neutrinos~\cite{Birrell:2012gg}. 
 
In chemical equilibrium the ratio of bottom quark (pair, $b$, $\bar b$) density $n_b^{th}$ to entropy density $\sigma=S/V$ just above quark-gluon hadronization temperature $T_\mathrm{H}\simeq150\sim160\,\mathrm{MeV}$ is $n_b^{th}/\sigma=10^{-10}\sim10^{-13}$, see Fig.~\ref{number_entropy_b002} We discuss how these results arise in appendix. 

Considering  the $n_b^{th}/\sigma$ value,  there is sufficient abundance of $b$, $\bar b$ quarks  for  the proposed matter genesis mechanism to be relevant. This is true even if $b$, $\bar b$ quarks  disappear from particle inventory below $T_\mathrm{H}$. In Fig.~\ref{number_entropy_b002} we see  that the charm (quark pair $c$,$\bar c$) abundance at $T_\mathrm{H}\simeq150$--$160\,\mathrm{MeV}$ is $\sim\!\!10,000$ times greater: $b$,$\bar b$ quarks are embedded in a background comprising all lighter $u,d,s,c$ quarks and antiquarks, as well as gluons\;$g$.

%~~~~~~~Figure1~~~~~~~~~~~~~~~~~~~~~~~~~~~~~~~~~~~~~~~~~~~~~~~~~~~~~~~~~~~~~~~~~~~~~~~~~~~~~~~~~
\begin{figure}[t]
\begin{center}
\includegraphics[width=3.5in]{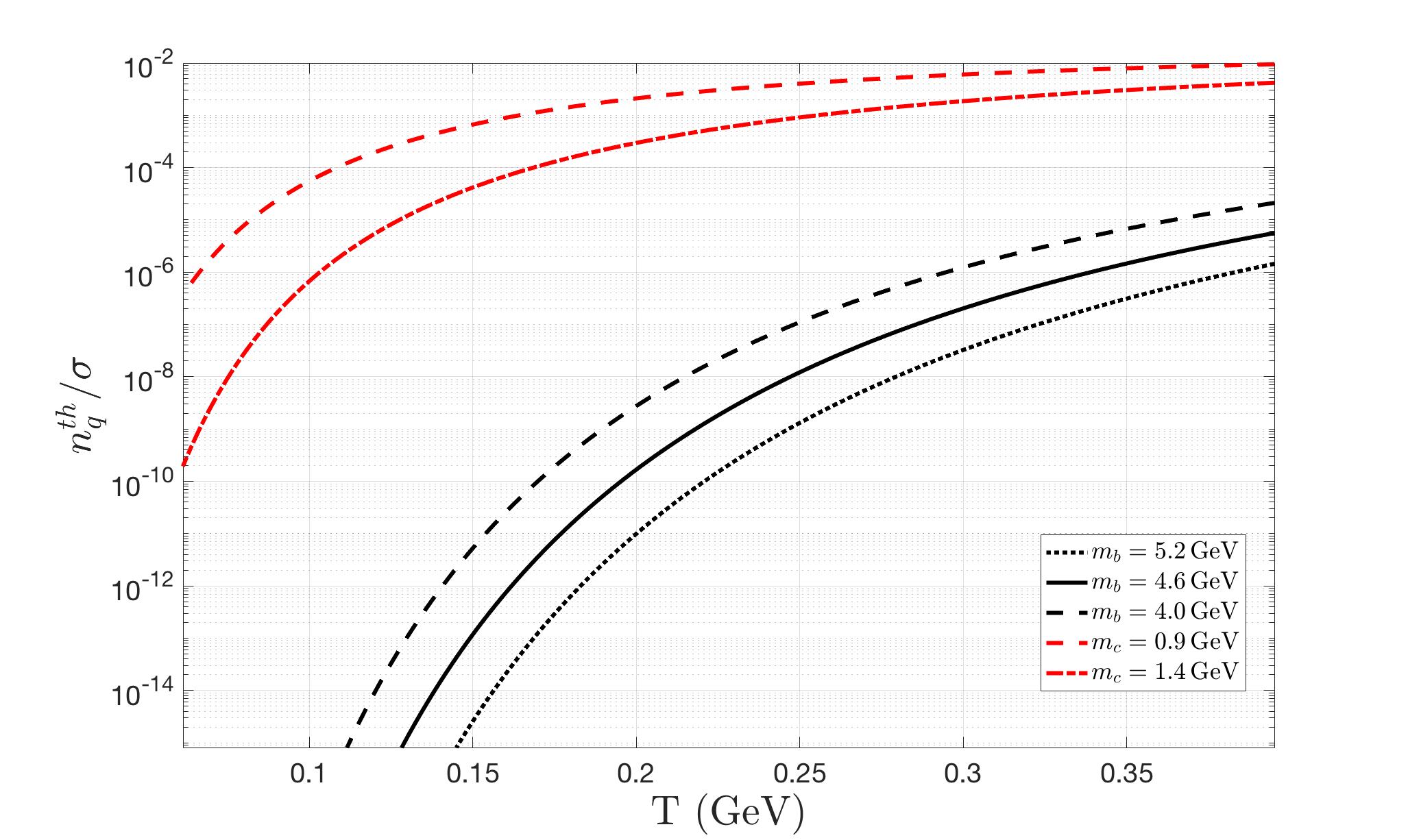}
\caption{The bottom  $b$,$\bar b$  and charm  $c$,$\bar c$ (pair) number density normalized by entropy density, as a function of temperature in the primordial Universe: $b$-quark mass parameters shown are $m_b=5.2\,\mathrm{GeV}$ (black dotted line), $m_b=4.6\,\mathrm{GeV}$(black solid line), and $m_b=4.0\,\mathrm{GeV}$(black dashed line). For  $c$-quark mass: $m_c=0.9\,\mathrm{GeV}$ (red dashed line) and $m_b=1.4\,\mathrm{GeV}$ (red dash-dotted line)}
\label{number_entropy_b002}
\end{center}
\end{figure}
%~~~~~~~~~~~~~~~~~~~~~~~~~~~~~~~~~~~~~~~~~~~~~~~~~~~~~~~~~~~~~~~~~~~~~~~~~~~~~~~~~~~~~~~~~~~~~

%%%%%%%%%%%%%%%%%%%%%%%%%%%%%%%%%%%%%%%%%%%%%%%%%%%%%%%%%%%%%%%%%%%%%%%%%%%%%%%%%%%
\section{Bottom quark freeze-out process}  
\subsection{Bottom production and annihilation}
Bottom quark freeze-out process occurs near to the QGP phase transition. This is so since in the expanding Universe at ever decreasing temperature the strong interaction gluon  $g+g\to b+\bar b$, and quark pair $q+\bar q \to b+\bar b$ fusion processes become  slower compared to the relatively slow WI decay process of bottom flavor: the lifespan of the preformed $\mathrm{B}_x$ meson states in empty-space has a $0.51 \sim 1.64$ picosecond lifespan~\cite{Tanabashi:2018oca}: 
\begin{align}\label{lifespan}
\tau_{\mathrm{B}_c^\pm}&=0.51\times\; 10^{-12}\;\mathrm{s}\;,\quad
\tau_{\mathrm{B}_s^0}=&\!\!\!\!\!1.51\times\; 10^{-12}\;\mathrm{s}\;,\\ 
\nonumber
\tau_{\mathrm{B}_d^0}&=1.52 \times\;10^{-12}\;\mathrm{s}\;,\quad
\tau_{\mathrm{B}_u^\pm}=&\!\!\!\!1.64\times\; 10^{-12}\;\mathrm{s}\;.
\end{align} 
Considering the energy balance in quark binding, and quark exchange reactions, we recognize that ultimately B$_c(b\bar c)$ mesons are always formed. The large binding of heavy ${\mathrm{B}_c^\pm}$ and its slow thermal motion protects this state, see also Ref.\;\cite{Schroedter:2000ek,Thews:1999bj}. In the following we assume that due to the enhanced binding effect of $\mathrm{B}_c^\pm$ all bottom $b$, $\bar b$ quarks are found in $\mathrm{B}_c^\pm$.

\begin{figure}[t]
\begin{center}
\includegraphics[width=3.5in]{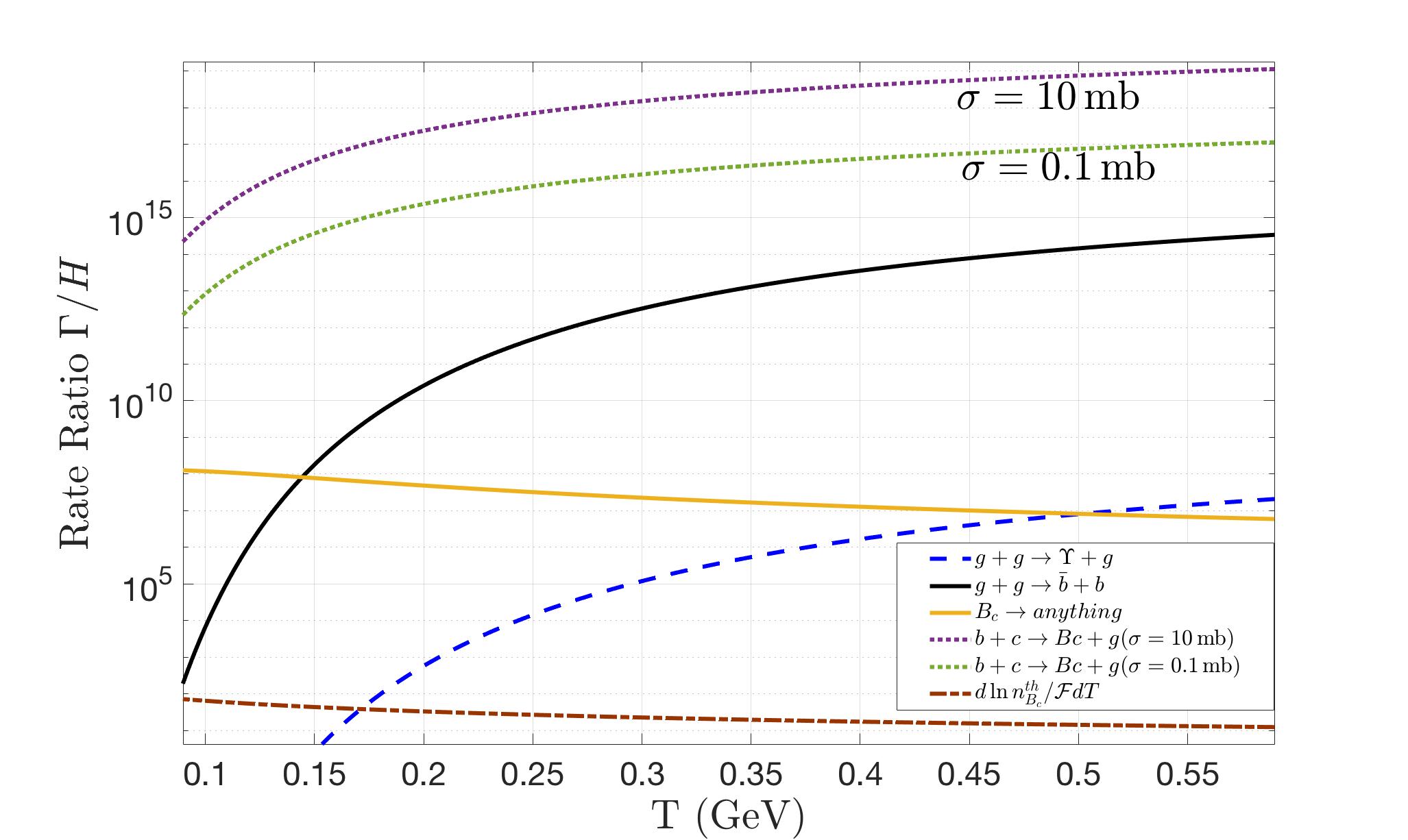}
\caption{The ratio between  reaction rates involving $\mathrm{B}_c^\pm$ meson and Hubble expansion rate $H$ as a function of temperature $T$ in the primordial Universe: $b+c\rightarrow \mathrm{B}_c+g$ (purple and green dotted line at the top with $\sigma\in 0.1,10$\;mb), $g+g\rightarrow b+\bar{b}$ (black solid line, mid figure), $g+g\rightarrow\Upsilon+g$ (blue dashed line), $\mathrm{B}_c\rightarrow\mathrm{anythings}$ (yellow solid line, horizontal mid-figure), and the density dilution term of $\mathrm{B}_c^\pm$ mesons (see text, brown dot-dashed line at the very bottom)}
\label{reaction_ratio}
\end{center}
\end{figure}

As noted, by means of quark exchange  through more abundant light quark states $B_x,x=u,d,s$, ultimately the most bound $B_c$ state arises with picosecond lifespan. The rapid formation rate of B$_c(b\bar c)$ states in primordial plasma is shown by dotted lines in Fig.~\ref{reaction_ratio}. We believe that this process is fast enough to allow consideration of bottom decay from the B$_c(b\bar c)$, $\overline{\mathrm{B}}_c(\bar b c)$ states. 

While B$_c$ is strongly bound and thus protected from follow-up chemical reactions, we need further to be sure that only a small fraction of produced $b,\;\bar b$ pairs forms the rapidly annihilating $b\bar b$-onium state such as $\Upsilon(b\bar b)$. An example could be the process
\begin{align}
\label{chembalancebb}
\mathrm{B}_c+\overline{\mathrm{B}}_c \Leftrightarrow b\bar b+c\bar c\;, \qquad Q\simeq 0\;. 
\end{align}
In a study of chemical process in QGP  by Yao and M\"uller~\cite{Yao:2017fuc}, their relative yield $\Upsilon/b\simeq 10^{-4}$ without charm. Using \req{BoltzN} and \req{FermiN}  we find the relative  charm abundance $c/(u+d+s) \approx 0.0015$ at $T_\mathrm{H}(m_c=1.24)$\;GeV. This means that charm catalyzed $\Upsilon(b\bar b)$ formation is sufficiently small not to alter our results in qualitative manner.

%~~~~~~~Figure2~~~~~~~~~~~~~~~~~~~~~~~~~~~~~~~~~~~~~~~~~~~~~~~~~~~~~~~~~~~~~~~~~~~~~~~~~~~~~~~~~
%~~~~~~~~~~~~~~~~~~~~~~~~~~~~~~~~~~~~~~~~~~~~~~~~~~~~~~~~~~~~~~~~~~~~~~~~~~~~~~~~~~~~~~~~~~~~~

The B$_c(b\bar c)$ lifespan is controlled by weak interaction processes and we will use the free space $\tau_{\mathrm{B}_c}=10^{-12}$s (nearly horizontal solid line in Fig.~\ref{reaction_ratio}). In comparison, the speed of Universe expansion with characteristic time at the scale of 10\;$\mu$s is nearly $10^{7}$ times slower, as seen in Fig.~\ref{reaction_ratio} where the ratio of rates to the Hubble expansion rate $H$ is shown. In Fig.~\ref{reaction_ratio} we further see that the lines depicting the rate of heavy flavor production   and  decay cross near to $T_\mathrm{H}\simeq150\,\mathrm{MeV}$. This result is of pivotal importance in this work as it establishes the temperature era for the abundance non-equilibrium of bottom quarks.

%%%%%%%%%%%%%%%%%%%%%%%%%%%%%%%%%%%%%%%%%%%%%%%%%%%%%
\subsection{Abundance nonequilibrium} 
\begin{figure}[t]
\begin{center}
\includegraphics[width=3.5in]{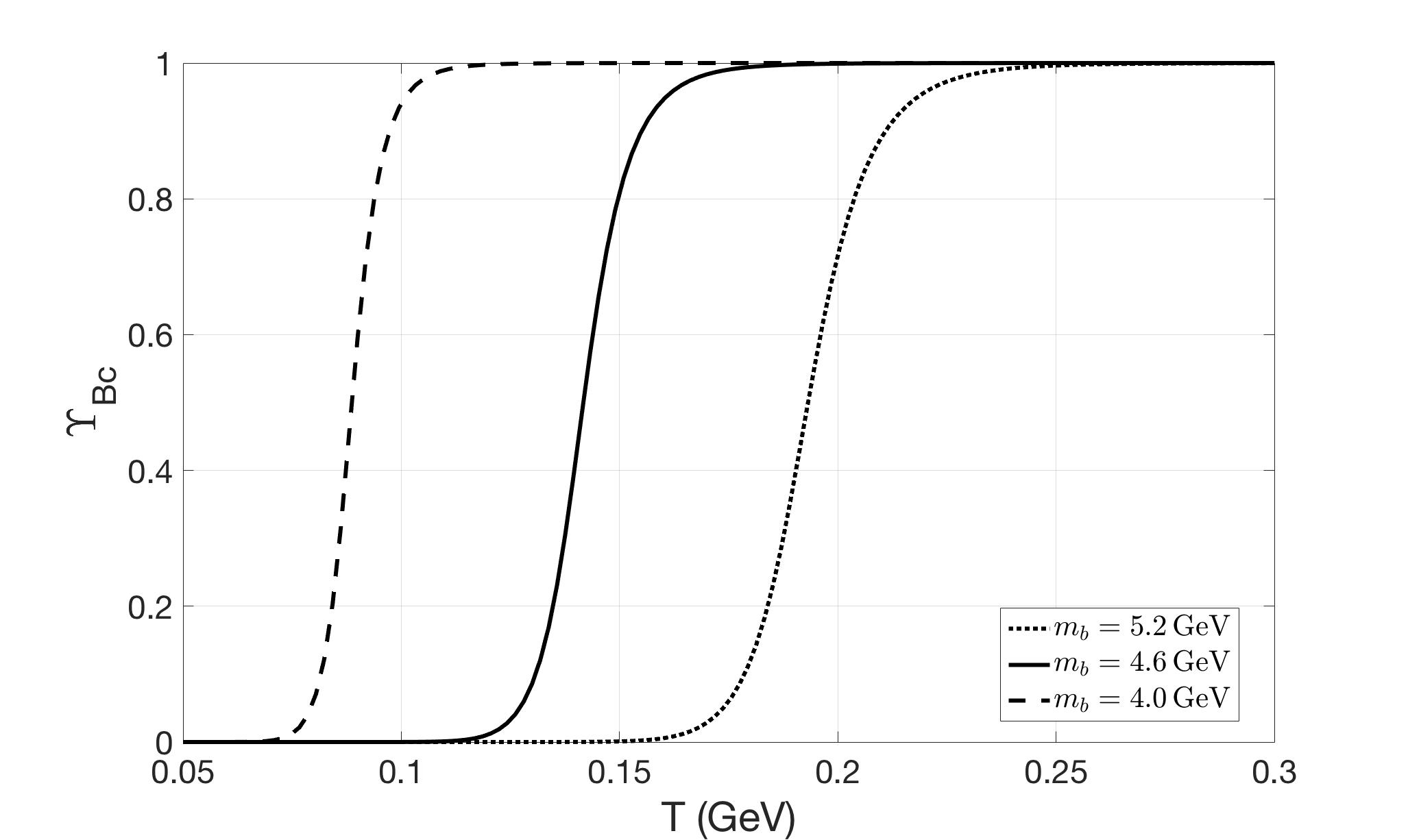}
\caption{The fugacity of $\mathrm{B}_c^\pm$ meson as a function of temperature in early Universe with different mass of bottom quark. We have $m_b=5.2\,\mathrm{GeV}$(dash-dotted line), $m_b=4.6\,\mathrm{GeV}$(solid line), and $m_b=4.0\,\mathrm{GeV}$(dashed line).}
\label{fugacity_bc}
\end{center}
\end{figure}

We obtain the bottom quark abundance fugacity by evaluating $\mathrm{B}_c$ fugacity as shown in Fig.~\ref{fugacity_bc} for several bottom quark masses entering the QGP gluon-fusion bottom pair production process. For bottom mass below 4.6\,GeV our results are also representative of the competition between strong interaction B$_c(b\bar c)$ forming processes in predominantly HG phase competing with WI decay process. 

The results seen in Fig.~\ref{fugacity_bc} are obtained by neglecting the time dependence of the fugacity, that is called adiabatic approximation -- we show in appendix that this approach is valid. The curves are thus analytical functions
\begin{align}
\label{Fugacity_Sol}
\Upsilon_{\mathrm{B}_c}=\frac{\Gamma^{\mathrm{Decay}}_{\mathrm{B}_c}}{2\,\Gamma^{\mathrm{Source}}_{\mathrm{B}_c}}\left[\sqrt{1+\left({2\Gamma^{\mathrm{Source}}_{\mathrm{B}_c}}/{\Gamma^{\mathrm{Decay}}_{\mathrm{B}_c}}\right)^2}-1\right]\;,
\end{align}
with characteristic rates of approach to chemical equilibrium -- the so called relaxation rates 
\begin{align}
\label{relaxation_time}
&\Gamma_{\mathrm{B}_c}^{\mathrm{Source}}\equiv\frac{R^{\mathrm{Source}}_{\mathrm{B}_c}}{n^{th}_{\mathrm{B}_c}}=\frac{R_{gg\rightarrow b\bar{b}}}{{n^{th}_{\mathrm{B}_c}}}\;,\quad
&\Gamma_{\mathrm{B}_c}^{\mathrm{Decay}}\equiv\frac{1}{\tau^{0}_{\mathrm{B}_c}}\;.
\end{align}
Here $R_{gg\rightarrow b\bar{b}}$, \req{Fusion_bb}, is the thermal reaction rate per volume of $g+g\longrightarrow b+\bar{b}$, and $n_{\mathrm{B}_c}^{th}$ is the thermal equilibrium number density of $\mathrm{B}_c^\pm$ mesons, obtained according to \req{BoltzN}.

The key result seen in Fig.~\ref{fugacity_bc} is that the large mass of bottom quark slows the strong interaction formation rate to the value of weak interaction B$_x$ decays just near the phase transformation of QGP to HG phase, were a mixed phase governs the transformation of phases lasting as long as $10\,\mu$s~\cite{Borsanyi:2016ksw,Fromerth:2012fe,Letessier:2002gp}.

%~~~~~~~Figure3~~~~~~~~~~~~~~~~~~~~~~~~~~~~~~~~~~~~~~~~~~~~~~~~~~~~~~~~~~~~~~~~~~~~~~~~~~~~~~~~~
%~~~~~~~~~~~~~~~~~~~~~~~~~~~~~~~~~~~~~~~~~~~~~~~~~~~~~~~~~~~~~~~~~~~~~~~~~~~~~~~~~~~~~~~~~~~~~~~

%%%%%%%%%%%%%%%%%%%%%%%%%%%%%%%%%%%%%%%%%%%%%%%%%%%%%%%%%%%%%%%%%%%%%%
\section{Possibility of bottom-catalyzed matter genesis}
\subsection{Repetitive formation and annihilation of bottom flavor} 
We can consider the duration of the phase transformation from the QGP to a hot hadronic gas to last in primordial Universe $\tau_\mathrm{H}\simeq10\,\mu s$. In this scenario, the number of bottom formation and annihilation cycling can be written as
\begin{align}
\mathrm{C_{ycle}}=\frac{\tau_\mathrm{H}}{\tau^0_{\mathrm{B}_c}}\approx2\times10^7\;.
\end{align}
Inspecting Fig.~\ref{number_entropy_b002} at the phase transition temperature $T=150\,\mathrm{MeV}$, we see the particle number per entropy for bottom quark $n_{b}^{th}/\sigma=n_b/\sigma\sim1.4\times10^{-13}$. To obtain the observed value of baryon asymmetry today we need to derive from each cycle an asymmetric excess of baryons over antibaryons
\begin{align}
\label{Basymmetry}
\epsilon=\left(\frac{B}{S}\right)\, \left(\frac{\sigma}{n_{b}}\right)\frac{1}{\mathrm{C_{ycle}}}\approx2.75\times10^{-5}.
\end{align}
To conclude, a small violation of matter-antimatter symmetry at the level of $\epsilon\sim10^{-5}$ associated with bottom quarks cycling during the hadronization can result in the observed baryon number today.

\subsection{How unique is the role of bottom flavor?} 
It is fortuitous that the here explored bottom freeze-out process happens near to the QGP cross-over to HG, $T_\mathrm{H}$. This coincidence is, however, is not a required condition for our non-equilibrium mechanism to operate, as already noted, our results can be refined by exploring bottomnium in HG environment. Since charm flavor is much lighter, a similar computation reveals charm flavor disappearance at considerably lower temperature $T\approx10\;\mathrm{MeV}$, thus well in the hadron phase and after most baryons and antibaryons emerging in QGP hadronization have annihilated~\cite{Fromerth:2002wb}. At this temperature the residual charm would be bound in D-mesons. 

Nothing protects the D-mesons from fusing into fast decaying charmonia. Qualitative studies we carried out lead us to believe that charm disappears sufficiently rapidly from particle inventory  once HG phase is formed and cannot contribute decisively to matter genesis. We have performed a similar study of strangeness flavor in great detail. Given the slower expansion of the Universe, we know with certainty that strangeness never occurs outside of full thermal equilibrium. We will return to this matter under separate cover.

We next look at the top quark flavor: If the top quarks freeze-out at higher temperature, the non-equilibrium conditions required for matter genesis will also operate. What stops a repetition of the here reported mechanism of matter genesis is the fast decay $t\to W+b$, with $\Gamma_t=1.4\pm0.2$\;GeV -- this also means that no bound state of top have time to form. Given the large value of $\Gamma_t$ we realize that top in hot QGP is produced in the $ W+b\to t$ fusion process -- given the strength of this process there is no freeze-out of the top quark till $W$ itself freezes out. However $W$ has an even greater width $\Gamma_W=2.1$\,GeV (and we note $\Gamma_Z=2.5$\,GeV); thus, $W, Z$ never freeze out, just like $\pi^0$~\cite{Kuznetsova:2008jt}. We conclude that $t$, $W$, $Z$ disappear gradually, retaining their thermal equilibrium abundance during the cooling of the primordial Universe.

\section{Conclusions}
We have  demonstrated the special role of the bottom flavor for matter genesis in primordial-QGP hadronization. The bottom quark flavor is just at the \lq sweet-spot\rq\ allowing the two Sakharov conditions in the same range of temperature, near to $T_\mathrm{H}$, the era of QGP hadronization. For those believing in the Anthropic principle the here proposed bottom flavor role in matter genesis provides beyond CKM-phase another reason for the existence of the third quark family.

Bottom flavor disappearance from particle inventory defines the epoch of required departure from equilibrium. Formation of a matter excess over antimatter at a relatively late Universe evolution period has the additional merit of depending on experimentally accessible Universe evolution stage: Further insight can be derived from  experimental study of bottom flavor in RHI collisions, and  more generally, exploration of the  the properties of bottom flavored particles, including but not limited to the search for a specific baryon non-conservation mechanism. 

Our results provide as we believe a strong  motivation to explore physics of baryon nonconservation involving the bottomnium mesons or/and bottom quarks in thermal environment. We have shown that millions of decays and reformation processes of bottom flavored quarks during the hadronization era can occur. This circular Urca-like situation can amplify even a tiny value of matter conservation violation producing the today observed matter-antimatter asymmetry.

\begin{acknowledgements} We thank Berndt Mueller for reading the manuscript and kind comments and suggestions.
\end{acknowledgements}

%%%%%%%%%%%%%%%%%%%%%%%%%%%%%%%%%%%%%%%%%%%%%%%%%%%%%%%%%%%%%%%%
 
 \appendix*
\section{Technical details} 
\subsection{Bottom flavor equilibrium abundance} 
In the primordial-QGP Universe evolution era, at $T\gtrapprox T_\mathrm{H}\simeq150\,\mathrm{MeV}$ freely propagating light quarks $u,d,s$ and gluons dominate the strongly interacting plasma in the Universe. The dominant mechanism for producing quarks is the gluon-fusion reaction $gg\rightarrow q\bar{q}$, hence the abundance of quarks is strongly coupled to gluons, both follow thermal equilibrium. The thermal equilibrium number density of heavy $Q=b,c$ quarks near to $T_\mathrm{H}$ with $m_{b,c}/T_\mathrm{H}\gg 1$ can be well described by the first term in the Boltzmann expansion of the Fermi distribution function 
\begin{subequations}
\begin{align}\label{BoltzN}
n_{Q}^{th}=\frac{g_{q}}{2\pi^2}\,T^3\sum_{n=1}^{n=\infty}(-1)^{n+1}\frac{\Upsilon^n}{n^4}\left(\frac{n\,m_{q}}{T}\right)^2\,K_2(n\,m_{q}/T)\;.
\end{align} 
For $s$ quarks several terms are needed. For light quarks we have to evaluate the massless limit
\begin{align}\label{FermiN}
n_q^{th}=\frac{g_{q}}{2\pi^2}\,T^3 F(\Upsilon)\;, \quad F=\int_0^\infty \frac{x^2dx}{1+\Upsilon^{-1}e^x}\;,
\end{align} 
\end{subequations}
where $ F(\Upsilon=1)=3\,\zeta(3)/2$ with the Riemann zeta function $\zeta(3)\approx1.202$.

The entropy density $\sigma$ of the early Universe is given by
\begin{align}\label{entroU}
\sigma=\frac{2\pi^2}{45}\,g_\ast^s\,T^3,
\end{align}
where $g_\ast^s$ counts the total number of effective degrees of freedom from entropy~\cite{Fromerth:2002wb,Fromerth:2012fe,Kolb:1990vq}. Near $T_\mathrm{H}$ only light particles matter in establishing the value of $g_\ast^s$; thus the result we consider is independent of actual abundance of $c, b$ and other heavy particles.

In Fig.~\ref{number_entropy_b002} the ratio of \req{BoltzN} with \req{entroU} is seen, the equilibrium number density per entropy density of heavy quarks. We evaluated this ratio for the bottom quark as a function of temperature $T$, allowing for different mass $m_b=4.0,\;4.6,\;5.2$\,GeV. The larger heavy quark mass should be used in the context of low energy processes we consider at $T\gtrapprox T_\mathrm{H}$, the lower values apply at higher energy scale~\cite{Tanabashi:2018oca}. The vale $m_b\simeq 5.2\,\mathrm{GeV}$ is used as potential model mass in modeling bound states and $m_b=4.0,\,4.6\,\mathrm{GeV}$ is the current quark mass at low and high energy scale. 

We note that naively the thermal abundance of $\mathrm{B}_c$ mesons of mass $m_{\mathrm{B}_c}=6.275\;\mathrm{GeV}$ should be smaller. However, such an abundance has to be computed allowing for a free supply of charmed quarks and antiquarks. This introduces a chemical potential that effectively reduces $m_{\mathrm{B}_c}$ by charm mass. This enhances the $\mathrm{B}_c$ meson yield to be in the realm of where the here presented results for unbound bottom quarks are shown. The same applies to other bottom quark preformed states in QGP.

%%%%%%%%%%%%%%%%%%%%%%%%%%%%%%%%%%%%%%%%%%%%%%%%%%%%%%%%%%%%%%%%
\subsection{Reaction rates involving bottom quarks}
The thermal reaction rate per volume for bottom quark production can be written as~\cite{Letessier:2002gp}
\begin{align}
\label{Fusion_bb}
R_{gg\rightarrow b\bar{b}}=\int^\infty_{s_{th}}ds\,\frac{dR_{gg\rightarrow b\bar{b}}}{ds}=\int^\infty_{s_{th}}ds\,\sigma_{gg\rightarrow b\bar{b}}\,P_g,
\end{align}
where $\sigma_{gg\rightarrow b\bar{b}}(s)$ is the cross section of the reaction channel $gg\rightarrow b\bar{b}$ and $P_g(s)$ is the number of collisions per unit time and volume. The cross section of gluon fusion is given by 
\begin{align}
\sigma_{gg\rightarrow b\bar{b}}=\frac{\pi\alpha_s^2}{3s}\bigg[&\left(1+\frac{4m^2_b}{s}+\frac{m^4_b}{s^2}\right)\ln{\left(\frac{1+w(s)}{1-w(s)}\right)}\notag\\
&-\left(\frac{7}{4}+\frac{31m^2_b}{4s}\right)\,w(s)\bigg],
\end{align} 
where the function $w(s)\equiv\sqrt{1-{4m^2_b}/{s}}$, and $m_b$ is the mass of bottom quark, $\alpha_s$ is the QCD coupling constant. 

The number of collisions per unit time and volume for massless gluons fusions into bottom quarks is given by
\begin{align}
P_g(s)=\frac{4T}{\pi^4}(\sqrt{s})^3\sum_{l,n=1}^\infty\,\frac{K_1(\sqrt{lns}/T)}{\sqrt{ln}}
\end{align}
Hence from \req{Fusion_bb} we can calculate the thermal production rate per volume for the gluon fusion as a function of temperature with given parameters. 

The $\alpha_s$ value we consider is based on required gluon collisions above $b+\bar b$ energy threshold; we adopt $\alpha_s=0.185$. The rate of formation normalized with the Hubble parameter is shown as a solid red line crossing the middle of Fig.~\ref{reaction_ratio}. The Hubble parameter is given by
\begin{align}
H^2=\frac{8\pi G}{3}\left(\rho_R+\rho_{SI }\right),
\end{align}
where $G$ is the Newtonian constant of gravitation, $\rho_R$ is the energy density of relativistic species, and $\rho_{SI}$ is the energy density from strong interaction in the early Universe~\cite{Letessier:2002gp}.

After formation, the heavy $b, \bar b$ quark can bind with any of the available lighter quarks, with the most likely outcome being a chain of reactions 
\begin{align}
b+q\to \mathrm{B}+g\;,\quad B+s\to \mathrm{B}_s+q\;,\quad \mathrm{B}_s+c\to \mathrm{B}_c+s\;,
\end{align}
with each step providing a gain in binding energy and reduced speed due to the diminishing abundance of heavier quarks $s, c$. To capture the lower limit of the rate of $\mathrm{B}_c$ production we show in Fig.~\ref{reaction_ratio} the expected formation rate by considering the direct process $b+c\rightarrow \mathrm{B}_c+g$, considering the range of cross section $\sigma=0.1\sim10\,\mathrm{mb}$. We obtain
\begin{align}
\Gamma(b+c\rightarrow \mathrm{B}_c+g)\approx H\times (10^{16}\sim10^{14})\;.
\end{align}

Despite the low abundance of charm, the rate of $\mathrm{B}_c$ formation is relatively fast, and that of lighter flavored B-mesons is substantially higher. Note that as long as we have bottom quarks made in gluon fusion bound practically immediately with any quarks $u, d, s$ into B-mesons, we can use the production rate of $b, \bar b$ pairs as the rate of B-meson formation in the primordial-QGP, which all decay with lifespan of pico-seconds, see \req{lifespan}. In Fig.~\ref{reaction_ratio} the mid-figure horizontal line shows $\mathrm{B}_c$ decay rate normalized with the Hubble parameter.

In Fig.~\ref{reaction_ratio} we also show the rate of direct production of Upsilonium (dashed blue line), which we obtain using Upsilonium three gluon lifetime $\tau^0_\Upsilon$ and unitarity of the reaction matrix element. We have
\begin{align}
R_{gg\rightarrow\Upsilon g}=\frac{2g_g}{(2\pi)^2g_\Upsilon}\frac{T}{m_\Upsilon\,\tau^0_\Upsilon}\int^\infty_{m_\Upsilon^2}ds\frac{s-m^2_\Upsilon}{\sqrt{s}}K_1(\sqrt{s}/T),
\end{align}
where $m_\Upsilon=9.460\,\mathrm{GeV}$ is the mass of Upsilonium. This rate is 7 orders of magnitude smaller compared to the direct production of open flavor. This is so since near temperature $T_\mathrm{H}$ it is hard to find two gluons above required energy threshold.

%%%%%%%%%%%%%%%%%%%%%%%%%%%%%%%%%%%%%%%%%%%%%%%%%%%%%%%%%%%%%%%%
\subsection{Bottom quarks abundance in dynamic model}
We consider the production and decay reaction processes based on the hypothesis that all bottom flavor is bound rapidly into $\mathrm{B}_c^\pm$ mesons. The master equation has two components, as we can jump into $\mathrm{B}_c^\pm$ meson states from $b$-pairs produced in gluon fusion reactions
\begin{align}
\label{Bc_source}
g+g\longleftrightarrow b+\bar b\;[b(\bar{b})+\bar{c}(c)]\longleftrightarrow &\mathrm{B}_c^\pm+g,\\
\label{Bc_decay}
 \mathrm{B}_c^\pm\longrightarrow &\mathrm{anything}.
\end{align}
For reactions \req{Bc_source} and \req{Bc_decay}, the master equation can be written as: 
\begin{align}
\label{Bc_eq}
\frac{1}{V}\frac{dN_{\mathrm{B}_c}}{dt}=\big(\,1-\Upsilon^2_{\mathrm{B}_c}\,\big)\,R^{\mathrm{Source}}_{\mathrm{B}_c}-\Upsilon_{\mathrm{B}_c}\,R^{\mathrm{Decay}}_{\mathrm{B}_c}\;,
\end{align}
where $R^{\mathrm{Source}}_{\mathrm{B}_c}$ and $R^{\mathrm{Decay}}_{\mathrm{B}_c}$ are the thermal reaction rate per volume of production and decay of $\mathrm{B}_c^\pm$ meson respectively. The bottom source rate is the gluon fusion rate \req{Fusion_bb}, while the decay rate is the natural lifespan, \req{lifespan} weighted with density of particles, see \req{BoltzN}
\begin{align}
R^{\mathrm{Decay}}_{\mathrm{B}_c}=\frac{n^{th}_{\mathrm{B}_c}}{\tau^{0}_{\mathrm{B}_c}}=n^{th}_{\mathrm{B}_c}\Gamma_{\mathrm{B}_c}^{\mathrm{Decay}}\;.
\end{align}

We wish to replace the variation of particle abundance seen on LHS in \req{Bc_eq} by the time variation of abundance fugacity $\Upsilon$. Considering the expansion of Universe we have
\begin{align}
\label{number_dilution}
\frac{1}{V}\frac{dN_{\mathrm{B}_c}}{dt}=\frac{dn_{\mathrm{B}_c}}{d\Upsilon_{\mathrm{B}_c}}\frac{d\Upsilon_{\mathrm{B}_c}}{dT}\frac{dT}{dt}+\frac{dn_{\mathrm{B}_c}}{dT}\frac{dT}{dt}+3Hn_{\mathrm{B}_c}\;.
\end{align}
The expanding Universe  conserves entropy and hence the relation between temperature and cosmic time can be written as
\begin{align}
\label{dTdt}
&\frac{dT}{dt}=-\frac{HT}{1+\frac{T}{3g^s_\ast}\frac{dg^s_\ast}{dT}}\equiv-\frac{H}{\mathcal{F}},\\
&\mathcal{F}\equiv\frac{1}{T}\left(1+\frac{T}{3g^s_\ast}\frac{dg^s_\ast}{dT}\right).
\end{align}
where $g^s_\ast$ is the degree of freedom describing the entropy in Universe~\cite{Kolb:1990vq}. 

Substituting these relations into \req{Bc_eq} the fugacity equation becomes
\begin{align}\label{Fugacity_Eq0}
\frac{d\Upsilon_{\mathrm{B}_c}}{dT}=&(\Upsilon_{\mathrm{B}_c}^2-1)\mathcal{F}\,\left(\frac{\Gamma^{\mathrm{Source}}_{\mathrm{B}_c}}{H}\right)\notag\\
&+\Upsilon_{\mathrm{B}_c}\mathcal{F}\,\left(\frac{\Gamma^{\mathrm{Decay}}_{\mathrm{B}_c}}{H}+3-\frac{d\ln{(n_{\mathrm{B}_c}^{th})}}{\mathcal{F}\,dT}\right).
\end{align}
In Fig.~\ref{reaction_ratio} we show  that when temperature is near to $T_\mathrm{H}$, we have $\Gamma^{Source}_{D_s}/H\approx\Gamma^{Decay}_{D_s}/H\sim10^{8}$, which is much larger than the term $d\ln{n^{th}_{D_s}}/\mathcal{F}dT\sim\mathcal{O}(10)$. In this case, the last two terms in \req{Fugacity_Eq0} can be neglected.

%%%%%%%%%%%%%%%%%%%%%%%%%%%%%%%%%%%%%%%%%%%%%%%%%%%%%%%%%%%%%%%%
\subsection{Bottom quark disappearance -- quantitative results}
We verify that the last two terms in \req{Fugacity_Eq0} are negligible; thus we solve
\begin{align}
\label{Fugacity_Eq}
\frac{d\Upsilon_{\mathrm{B}_c}}{dT}=(\Upsilon_{\mathrm{B}_c}^2-1)\mathcal{F}\,\left(\frac{\Gamma^{\mathrm{Source}}_{\mathrm{B}_c}}{H}\right)+\Upsilon_{\mathrm{B}_c}\mathcal{F}\,\left(\frac{\Gamma^{\mathrm{Decay}}_{\mathrm{B}_c}}{H}\right).
\end{align}
Furthermore we can use the adiabatic solution for the fugacity equation setting $d\Upsilon_{\mathrm{B}_c}/dT=0$ and solve algebraically the equation for $\Upsilon_{\mathrm{B}_c}$. We so obtain \req{Fugacity_Sol}.

%~~~~~~~Figure~~~~~~~~~~~~~~~~~~~~~~~~~~~~~~~~~~~~~~~~~~~~~~~~~~~~~~~~~~~~~~~~~~~~~~~~~~~~~~~~~~~~~
\begin{figure}[b]
\begin{center}
\includegraphics[width=3.35in]{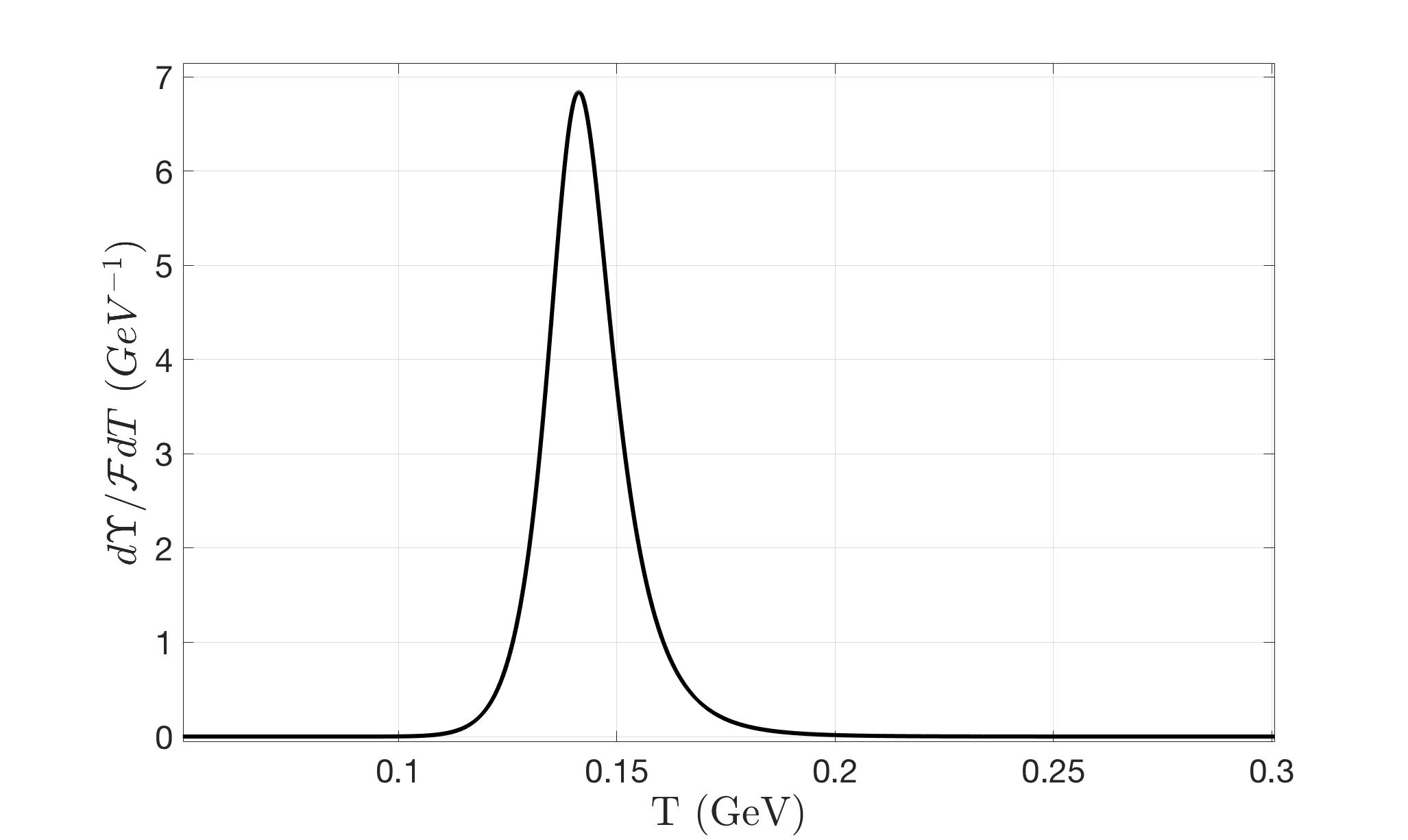}
\caption{The quantity $d\Upsilon/dT$ as a function of temperature in the Universe. We found that the values of $d\Upsilon/\mathcal{F}dT\ll\Gamma^{\mathrm{Source}}/{H},\,\Gamma^{\mathrm{Decay}}/{H}$.}
\label{dupsilon_dT}
\end{center}
\end{figure}
%~~~~~~~~~~~~~~~~~~~~~~~~~~~~~~~~~~~~~~~~~~~~~~~~~~~~~~~~~~~~~~~~~~~~~~~~~~~~~~~~~~~~~~~~~~~~~~~

In Fig.~\ref{fugacity_bc} the fugacity of $\mathrm{B}_c^\pm$ meson $\Upsilon_{\mathrm{B}_c}$ as a function of temperature, \req{Fugacity_Sol} is shown around the temperature $T=200\sim100\,\mathrm{MeV}$ for different masses of bottom quarks. In all cases we see prolonged non-equilibrium. This happens since the decay and reformation rates of bottom quarks are comparable to each other as we have noted in Fig.~\ref{reaction_ratio}, where both lines cross.

%%%%%%%%%%%%%%%%%%%%%%%%%%%%%%%%%%%%%%%%%%%%%%%%%%%%%%%%%%%%%%%%%%%%%%%%%%%%

\subsection{Validity of approximations} 
The results shown in Fig.~\ref{fugacity_bc} were obtained in adiabatic approximation neglecting the time dependence and dilution effects, which is possible since $H$ is small compared to process rates. We now establish that these approximations are justified.
In Fig.~\ref{dupsilon_dT} we show the quantity $d\Upsilon/\mathcal{F}dT$ for $\mathrm{B}_c$ meson as a function of temperature in the Universe. This result shows that the maximum value of $d\Upsilon/\mathcal{F}dT \sim\mathcal{O}(10)$ is more than $10^{-7}$ orders of magnitude smaller compared to the ratios ${\Gamma^{\mathrm{Source}}}/{H}$ and ${\Gamma^{\mathrm{Decay}}}/{H}$ and can be neglected. This establishes that the adiabatic solution is a good approximation for solving the $\Upsilon_{\mathrm{B}_c}$ in early Universe.

%%%%%%%%%%%%%%%%%%%%%%%%%%%%%%%%%%%%%%%%%%%%%%%%%%%%%%%%%%%%%%%%
%%%%%%%%%%%%%%%%%%%%%%%%%%%%%%%%%%%%%%%%%%%%%%%%%%%%%%%%%%%%%%%%

%%%%%%%%%%%%%%%%%%%%%%%%%%%%%%%%%%%%%%%%%%%%%%%%%%%%%%%%%%%%%%%%%%%
\end{document}